\documentstyle[aps,psfig]{revtex}

\def\ZZ{\hbox to 8.2222pt{\rm Z\hskip-4pt \rm Z}}
\newcommand{\al}{\alpha}

\newcommand{\de}{\delta}

\newcommand{\la}{\lambda}

\newcommand{\si}{\sigma}

\newcommand{\tree}{{\cal T}}

\newcommand{\Ga}{\Gamma}

\newcommand{\be}{\begin{equation}}
\newcommand{\ee}{\end{equation}}
\newcommand{\bqa}{\begin{eqnarray}}
\newcommand{\eqa}{\end{eqnarray}}
\newcommand{\ba}{\begin{array}}
\newcommand{\ea}{\end{array}}

\begin{document} 
\wideabs{
\title{A Rigorous Proof of Fermi Liquid
Behavior\\ for Jellium Two-Dimensional Interacting Fermions}
\author{M. Disertori and V. Rivasseau}
\address{Centre de Physique Th{\'e}orique, Ecole Polytechnique,
F-91128 Palaiseau Cedex, FRANCE}
\date{Received \today}
\tighten
\maketitle
 
\begin{abstract}
Using the method of continuous constructive renormalization group
around the Fermi surface, it is proved that a jellium two-dimensional 
interacting system of Fermions at low temperature $T$
remains analytic in the coupling constant  $\lambda$
for $|\lambda| |\log T| \le K $ where $K$ is some numerical 
constant and $T$ is the temperature. Furthermore in that range of
parameters, the first and second derivatives of the self-energy
remain bounded, a behavior which is that of Fermi liquids and in
particular excludes Luttinger liquid behavior. 
Our results prove also that in dimension two
any transition temperature must be non-perturbative in the coupling
constant, a result expected on physical grounds. 
The proof exploits the specific momentum conservation rules 
in two dimensions.
\end{abstract}
\draft
}

Conducting electrons in a metal at low temperature are well described
by Fermi liquid theory (FLT). However, it is known that the FLT is
not stable at strict zero temperature, even if the interaction is 
repulsive \cite{Kohn}. Indeed, below some critical temperature
the effective electrons or holes which are the excitations of the Fermi system
bound into Cooper pairs and the metal becomes superconducting.

With the discovery of the high temperature superconductors 
(HTSC), the Fermi liquid paradigm underlying the BCS theory has been 
questioned \cite{Anderson}. In fact, experiments showed that the physical 
properties of the normal phase of the HTSCs are in conflict with FLT 
(for a recent experimental survey, see Ref. \cite{Maple}). For this reason, 
new theoretical proposals have been done by different groups 
\cite{Theories}. In particular, a deeper understanding of 
FLT itself became necessary.      

During the last ten years a program has been designed to investigate
rigorously Fermionic  many-particle systems by means of field 
theoretical methods \cite{BG,FT,FMRT,S1}.
In particular the Wilson renormalization group 
has been extended to models with surface singularities such as the Fermi
surface. 
The ultimate goal is to create a mathematically rigorous
theory of the BCS transition and of similar phenomena in solid state physics.
This is a long and difficult program which requires to glue together several
ingredients, in particular renormalization group around the Fermi
surface and spontaneous symmetry breaking. 

In this paper we give a contribution towards a rigorous scenario for  
the description of the normal phase in Fermionic  many-particle systems.   
As we have already mentioned in the opening paragraph, FLT  
is not valid at zero temperature due to the 
superconducting instability. Even
when the dominant electron interaction is repulsive, the Kohn-Luttinger
instabilities \cite{Kohn} prevent the FLT to be generically valid
down to zero temperature. There are nevertheless two paths towards  a
mathematically rigorous FLT. The superconducting 
instabilities can be avoided by considering models in which the Fermi surface
is not invariant under $p \to -p$ \cite{FKLT}. 
In two dimensions it is possible to
prove (even non-perturbatively) that in this case the FLT
scenario remains valid at zero temperature.
However, this analysis requires a rigorous control of the stability of
a non-spherical Fermi surface under the renormalization group flow,
a difficult technical issue \cite{FST}.
The other alternative is  to study the FLT at finite temperature
above the superconducting 
transition temperature. A system of weakly interacting
fermions has an obviously stable thermodynamic limit
at high enough temperature, since 
the temperature acts as an infrared cutoff on the propagator
in the field theory description of the model. In this point of view,
advocated in Ref. \cite{S2}, the non-trivial theorem consists in showing
that stability (i.e. summability of perturbation theory) holds
for all temperatures $T$ satisfying to $|\lambda| |\log T| \le K$,
and that in that range of temperatures and couplings the first and second
derivatives of the self-energy with respect to the momenta
remain uniformly bounded.
This is what we have recently completed. We developed a rigorous continuous 
renormalization group method \cite{DR1}, and applied it to the simplest
two dimensional model of interacting Fermions, the jellium model \cite{DR2}.
(This model is a good physical approximation of the Hubbard model
in the regime of weak filling).  
In this way we proved a mathematically rigorous theorem,
stated below, which physically means
that the system above the BCS temperature is a Fermi liquid. Here we only
outline the proof of this theorem, and for more details we refer to \cite{DR2}.

Let $\hat{C}(k)$  be the Fermionic  propagator of the isotropic jellium
model at finite temperature $T$:
\be
\hat{C}_{ab} (k) = \de_{ab} \frac{\eta (k)}{ik_0-e(\vec{k})},
\quad \quad e(\vec{k})= \frac{\vec{k}^2}{2m}-\mu \ ,
\label{prop}
\ee
where $a,b \in \{1,2\}$ are the
spin indices and  $\eta (k)$ is some ultraviolet cutoff (the infrared cutoff
is given by the temperature). 
The vector $\vec k$ is two-dimensional.
The parameters $m$ and $\mu$ correspond to the effective mass and to
the renormalized 
chemical potential (which fixes the Fermi energy), and the Matsubara
frequencies $k_0$ take the discrete values
\be
k_0 =   \frac{2n+1}{\beta} \pi \ , \quad n \in \ZZ \ , \label{discretized}
\ee
where $\beta=1/T$, and we fixed $2m=1$.

Let the system of interacting Fermions be defined
by the Grassmann measure:
\be e^{S_V} d\mu_{C} (\bar \psi , \psi)
\label{measure}\ee
where $d\mu_{C} (\bar \psi , \psi)$ is the Grassmann Gaussian measure 
with covariance $\hat{C}$ given by Eq. (\ref{prop}) and 
the interaction in a finite volume $V$ is
\be
S_V = \frac{\la}{2} \int_V d^3x\; \left ( \sum_a \bar \psi_a\psi_a\right )^2
\ + \ \de \mu  \int_V d^3x\; \sum_a \bar \psi_a  \psi_a \ ,
\label{actionr}\ee
$\la$ being the bare coupling constant and $\de \mu$ being the bare 
chemical potential counterterm. The self-energy $\Sigma(k)$ is defined
as the sum over all non-trivial 1PI two point subgraphs (i.e those
not reduced to a single $\de \mu $ counterterm). 
The following theorem, advocated in \cite{S2} is proved in \cite{DR2}: 

\medskip\noindent{\bf Theorem.} \ {\it
The vertex functions (and the connected correlation functions) of the
measure (\ref{measure})
are analytic in the bare coupling constant $\la$, for $|\la| \le c$, with 
$c$ given by the equivalent relations
\be
T = e^{-{K \over c  }}\quad ; \quad   
c= {K \over |\log T | }
\ee
for some constant $K$  (this relation is limited to the
interesting low temperature regime $ T < 1$). Furthermore
in that range all the first  partial derivatives  of the
self energy on the Fermi surface
and all the second partial derivatives  of the
self energy anywhere in momentum space 
(in every direction) remain bounded:
\be \left | {\partial \over \partial k_{i} } \Sigma _{|{k_{0}={\pi\over\beta},
e(\vec k)=0}} \right | \le K_1 |\lambda| ^2 \label{first}
\ee 
\be\left |\left |
{\partial^{2}\over \partial k_{i}\partial k_{j} }
\Sigma (k) \right |  \right |_{\infty} \le K_2 \label{second}
\ee
where i and j take values 0,1,2, and $K_1$, $K_2$  are  some constants.
}

This theorem rules out Luttinger liquid behavior, since for a Luttinger liquid
the first derivative of $\Sigma$ would be bounded only
by $K_1 |\lambda|$ in the range considered, and the
second derivative of $\Sigma$ would behave as $\la^{2}/T$, which is
certainly not uniformly bounded for $|\la| |\log T| \le K$.  In 
fact in dimension $d=1$ this Luttinger
liquid behavior has been proved up to $T=0$ \cite{BGPS,BM}.
Remark that 
for the ``Fermi surface'' conditions we took $k_{0}={\pi\over\beta}$ since
we cannot take $k_{0}=0$ which would violate antiperiodicity of the Fermions.

In the following we will give an idea of the
mathematical techniques used to obtain this result, in the hope that
they might be of interest to a general audience of physicists. 

Fermionic  path integrals are based on the rules of noncommutative or
Berezin integration. 
This means that physical quantities
can be expressed in terms of determinants.
The correlation functions are then defined by
\bqa
&& S(y_{1},...,x_{p}, z_{1},...,z_{p})\nonumber\\
&&=\int d\mu_{C}(\psi, \bar \psi )
\prod_{i=1}^p \bar{\psi}_{a_i}(y_i) \psi_{b_i}(z_i)
 e^{S_{V}(\bar\psi, \psi)}  \nonumber\\
&& = \sum_{n,n'=0}^\infty 
   { \la^{n}\over  n!} {\de\mu^{n'}\over n'!}
\int_V dx_{1}...dx_{n+n'} \det(\{x_i\},\{y_i\},\{z_i\})\label{act}
\eqa
where  $d\mu_{C}$ is the Fermionic  Grassmann Gaussian measure
with propagator $C$ and the determinant 
$\det(\{x_i\},\{y_i\},\{z_i\})$ 
corresponds to the sum over all Wick contractions of the fields.

To compute intensive
quantities in the thermodynamic limit, one needs to sum over connected
graphs only. Now, the simplest structures connecting points are trees,
and each connected Feynman graph contains a tree. 
So why not reorganize the
series by summing over trees only, keeping the remaining loop structure
of the graph into the closed form of a determinant? This is the point of
view adopted in Ref. \cite{DR1}. It has the
advantage that since there are much less trees 
than graphs, and since determinants, by Gram's or Hadamard inequalities
can be easily bounded, the convergence of the Fermionic  series follows.
Two difficulties have to be solved.
The  first problem comes from the fact that there are typically
several trees in a Feynman graph, so which one should one pick? This is a
combinatoric problem which is completely solved by the invention
of ``tree formulas'' which express in the most symmetric way how picking
a tree affects the combinatoric of the remaining loop lines \cite{AR1}. 
The key
point is that these tree formulas preserve positivity, hence 
Gram's or Hadamard inequalities \cite{L,AR2}.
The connected amputated Schwinger functions, in the thermodynamic limit,
 may then be written as
\bqa
&&\Ga_{2p}(\phi_1,...\phi_{2p}) = 
 \sum_{n,n'=0}^\infty  { \la^{n}\over  n!} {\de\mu^{n'}\over n'!}\nonumber\\
&&\sum_{\tree\in T_{n+n'}} \int_0^1 \prod_{i=1}^{n+n'-1} dw_i
\int dx_{1}...dx_{n+n'}   \prod_{i=1}^{2p} \phi_i(x_i) \nonumber\\
&&\left [\prod_{l\in \tree}\; C_l\right ]\; 
\det_{\rm loop}(\{w_i\})
\label{connect} \eqa
where $\phi_1,...\phi_{2p}$ are some test functions we have inserted instead
of the amputated external propagators,  $T_{n+n'}$  
is the set of trees connecting the  $n+n'$
vertices (hence with $n+n'-1$ lines) , the parameter $w_i$
is associated by the tree formula to  the tree line propagator $C_{l_i}$,  
$\det_{\rm loop}(\{w_i\})$ denotes the determinant of the matrix   
 containing the uncontracted loop fields, whose propagators depend from
the $w_i$ parameters too. 
Without performing renormalization one can prove, for high 
enough temperature, that 
this sum is absolutely convergent. The strategy of the proof 
is the following. 

The determinant is bounded using a Gram inequality
which states that
\be
|\det <f_{i},g_{j}> | \le \prod_{i} || f_{i} || \prod_{j} || g_{j} || 
\ee
where $f_{i}$ and $g_{j}$ are any sets of $n$ vectors in an Hilbert space.
The  spatial decay of the tree line propagators is used to 
perform the  spatial integrals over $x_2$...$x_{n+n'}$, the first 
integral being performed with the help of the test function $\phi_1$;
the result of these operations  is some function $f(\tree, T)$.
The sum over trees is simply bounded by 
\be
\sum_\tree f(\tree)\leq \# T_{n+n'} \; \sup_\tree f(\tree).
\ee
Now, by the Cayley's theorem, 
the number of trees satisfies  $\# T_{n+n'}\leq  (n+n')^{n+n'} $
and is compensated by the factors $1/n!n'!$ in 
Eq.(\ref{connect}).  We obtain that the series
Eq.(\ref{connect}) is bounded by  
\be
\left |\Ga_{2p}(\phi_1,...\phi_{2p})\right |\leq \sum_{n,n'} (|\la| K_1)^n
(|\de\mu| K_2)^{n'},
\ee
where $K_1$ and $K_2$ depend on $T$. This sum is convergent 
for $\la$ and $\de\mu$ small enough, depending on $T$. 

The tree formalism then allows us to 
compute the thermodynamic limit of Fermionic systems with cutoffs 
easily in a transparent and convergent way. It is 
the right compromise between the unexpanded Berezin determinant which is good
for convergence of the series in a finite volume but unadapted
to thermodynamic limit, and the Feynman graphs which are good
for thermodynamic limit but too many for convergence of the series.

The second difficulty was raised in particular in Ref. \cite{S2}: 
is it possible to perform renormalization
and compute continuous renormalization group flows in such a 
tree formalism? This is not obvious because it is known that
renormalization
is due to the loop structure of Feynman graphs, so how to see this 
loop structure
if loop lines remains unexpanded in a determinant? The answer to
that question is twofold. 
First, the $w$ parameters in the tree formula can be combined with
a scale analysis to ensure that the tree which is picked in a graph is
always optimal in the sense that hard ``short-range'' propagators are
always selected in priority with respect to softer ones. As a result,
each $w_i$ parameter fixes the energy of the tree line $l_i$ 
and constrains loop lines connecting $T_i$ and $T'_i$ to be softer than $l_i$
(where $T_i$ and $T_i'$ are the two subtrees connected by $l_i$).
Then the energy parameters $w_i$  can be ordered, 
cutting the impulsion space in a set of bands.
Second, once this scale analysis has been performed, a partial
expansion of the loop determinant 
can detect all the dangerous two and four point functions
which require renormalization. A key point is that this 
expansion can be done without destroying the Gram bound, and the
corresponding sum is not too big (this means its cardinal remains bounded
by $K^n$   (where $K$ is a constant))  
because in typical graphs there are not many two and four point
subgraphs.

Let us see in more detail the technique.
The  renormalization group analysis of the theory is performed,
in momentum space, according to the distance from the Fermi surface
$k_{0}= \pi/\beta $, $e(\vec k)=0$. The propagator is decomposed 
over an auxiliary parameter $\al$, running from $1$ to $\infty$, such that
this distance is roughly $\alpha ^{-1/2}$. (This decomposition
is written as an integral to preserve the continuity of the renormalization
group flows).
The corresponding propagator
$C_{\alpha}$ decays on a range $\sqrt{\alpha}$. But this naive scale analysis
does not lead to the right power counting; it is necessary to perform
a second decomposition with respect to the angular direction of  $\vec k$. 
Combining this angular decomposition into ``sectors''
with the momentum conservation
rule in two dimensions we recover the correct power counting of a just 
renormalizable field theory \cite{FMRT}.

However some complications are worth being mentioned. The consideration
of almost collapsed vertices in which the four angular directions are almost
parallel forces us to adopt ``anisotropic sectors'' which have length 
$\alpha ^{-1/4}$ in the tangential direction, much longer
than their radial width $\alpha ^{-1/2}$.

Since a vertex is hooked to four different propagators
with possibly different scales, the angular decomposition
cannot be done just once, but must be updated progressively, taking also
into account the momentum conservation of larger and larger subgraphs
as the renormalization group flow progresses towards the Fermi surface.
This is one of the most delicate technical points in \cite{DR2}.

Once the correct power counting of a renormalizable theory has been obtained,
it remains to perform correctly the renormalization of the Fermi radius
induced by the interaction; indeed since the two
point subgraphs are relevant rather than marginal, our bound for the critical
temperature remains a power of $\la$ instead than exponentially small,
as long as this renormalization is not performed.

As we have already mentioned, it 
is possible through some auxiliary expansion of the loop determinant
to test for the
presence or absence of two point subgraphs to be renormalized, since this
auxiliary expansion does not generate any factorial \cite{DR1}. 
But a last difficulty has to be overcome. Constructive renormalization has
to be performed by subtracting in direct space, not in momentum space.
Let $G(x,y)$ be the amplitude of a two point subgraph to renormalize,
and $C_{\sigma}(.,x)C_{\sigma}(y,.)$ be its two external propagators, in an
angular sector $\sigma$.
In the subtraction process, the tree decay inside the two point subgraph
must be used to compensate for the gradient term
$\nabla -\si$ acting on the external
propagators of the subgraph. The tree decay inside $G$ is anisotropic,
and can happen in a different direction than the sector $\sigma$. In that case
a further auxiliary expansion has to be performed to take into
account the loop structure of $G(x,y)$; without that auxiliary expansion,
renormalization would not work as expected. This last difficulty was 
never recognized before.

The bound on the first derivatives of the self-energy (\ref{first})
is then an easy consequence of this additional loop analysis. 
The factor $\lambda^2$ in this bound already excludes the Luttinger
liquid behavior.
Finally to obtain that the second derivatives of the self-energy
$\Sigma$ remain bounded, hence to prove the Fermi
liquid behavior, we have to perform a second auxiliary analysis which refines
anisotropic sectors of one four point function into isotropic sectors.
This bound consumes a second logarithm because of
the possibility of collapse for this four point function, and this explains 
that there is no coupling constant left in the bound (\ref{second}).

In three dimensions such a detailed analysis with angular decomposition
of the propagator is no longer possible because the momentum conservation
rules are less rigid and allows for the possibility of non-planar
vertices. But using the Hadamard bound on determinants (which is worse
than the Gram bound in the sense that it consumes the factor $1/n!$ of
symmetry), we hope to prove the same exponentially
small bound on the BCS transition temperature. Essentially one should combine
the result of \cite{MR} 
with a multiscale cluster expansion, and the renormalization
of two-point functions.

The theorem in \cite{DR2}, combined with four point function 
renormalization and the infrared
analysis of the Goldstone boson, may also lead in the future
to a rigorous analysis
of the superconducting phase of two-dimensional
Fermi liquids. The analysis performed in \cite{DR2} should
also be generalized in the future to study other types of instabilities. 
Particularly interesting is the case of a nesting of the  Fermi surface 
in the Hubbard model at half-filling. Nesting in this case is 
responsible for the appearance of antiferromagnetic instabilities. A 
mathematically rigorous analysis of this case is however very 
difficult and is not simply a trivial extension of the analysis 
performed in this paper. The point is that the diamond shaped Fermi surface 
introduces new technical difficulties as, for instance, the presence 
of van Hove singularities and flat sides of the diamond 
\cite{FST}\cite{Metz}.  
\medskip 

We thank M. Salmhofer for explaining us the meaning and importance
of bounds (\ref{first}-\ref{second}) on the self-energy derivatives
in order to distinguish between Luttinger and Fermi liquid behavior. We also
thank F. Nogueira for discussions on the content of this letter.

\end{document}